\documentclass[twocolumn,aps,pre]{revtex4}
\usepackage{graphicx,color}
\begin{document}
\title{\bf{ Multi-shocks in asymmetric simple exclusions processes: 
Insights from fixed-point analysis of the boundary-layers}}
\author{Sutapa Mukherji}
\affiliation{Department of Physics, Indian Institute of Technology,
Kanpur-208 016}
\date{\today}
\begin{abstract}
The boundary-induced phase transitions in an 
 asymmetric simple exclusion process with inter-particle repulsion 
and bulk non-conservation are  analyzed through the fixed points 
of  the boundary layers. This system is known to have  
phases in which particle density profiles have different kinds of shocks.  
We show how this boundary-layer fixed-point 
method allows us to gain  physical insights on the 
nature of the phases and also to obtain several quantitative 
results on the density profiles especially on the nature of 
the boundary-layers and shocks.

\end{abstract}
\maketitle
\section{Introduction}
Asymmetric simple exclusion process (ASEP)
 is a non-equilibrium process in which 
particles hop to the neighboring site in a specific direction 
 following  certain conditions \cite{ligett}. 
In the simplest model, particles obey the  mutual exclusion condition 
due to which  a site  cannot be  
occupied by more than one particle and a hop to the neighboring  site 
 is possible provided this target   site is empty.  The particles 
after being injected
at one end of the lattice at a rate $\alpha$ hop across the lattice and 
reach the other end where they are withdrawn at a rate $1-\gamma$. 
There exist other models where   particles 
can have  attractive  or repulsive interaction in addition to the 
exclusion interaction \cite{kls,hager}.

All these systems exhibit 
interesting boundary-induced phase  transitions for which the tuning 
parameters are the boundary  rates, $\alpha$ and $\gamma$ 
\cite{krug,straley}. 
In different phases, the average particle density has 
 distinct  constant values across the bulk of 
the lattice.  These particle density profiles in different 
phases may also differ due to different location 
and nature of the boundary-layers. 
More features, such as   coexistence of high and low-density 
regimes  are seen in systems where particle 
number is not conserved in the bulk
due to  attachment and  detachment of particles to and   from the
lattice \cite{parameg,smsmb}.
In this coexistence phase (also known as a shock phase), 
the particle density profile  
has a jump discontinuity (shock)  in the interior of the lattice 
from a  low to a   high density  value.
The slope and  the number of such shocks in a density profile 
are related to the 
nature of the inter-particle interactions which determine 
the fundamental current density relation \cite{sm,rakos}.  
Although this relation   predicts  
the kind  of  shocks that can be seen \cite{rakos} 
in the density profile,  a 
systematic characterization of the phase-transition 
and the phase diagram 
in the  $\alpha-\gamma$ plane requires a detailed  analysis 
of the relevant equations describing the dynamics 
in the steady-state \cite{smb,smsmb,jaya}. 
Drawing analogies from the equilibrium phase 
transitions,  first order, critical \cite{straley,parameg}
 and tricritical \cite{jaya}
kind of phase 
transitions have been observed so far  in various ASEP models. 

While characterizing the phase transitions, 
 it is useful to study  the  variation of the 
height or the width of the boundary-layers  as $\alpha$ and $\gamma$ 
are changed.  
In a way, these boundary-layers play  important roles in deciding 
the "order parameter" like quantities in these 
non-equilibrium phase transitions. 
Owing to their importance in describing  
phase transitions,  the boundary-layers for several interacting and
 non-interacting models have been studied   
using the techniques of boundary-layer analysis \cite{cole}. 
It is found that  the system usually 
enters into a shock phase from a non-shock phase  
due to the   deconfinement of the 
boundary-layer from the boundary. This 
 deconfinement can be described by a nontrivial 
scaling exponent associated with the width of the boundary-layer \cite{smsmb}.  
For the pure exclusion case, except for a critical point, 
the  transition from a non-shock to a shock phase 
is first order in nature  since  a shock 
of finite height is  formed on the
phase boundary. The height of the shock on  the phase boundary  
reduces as one approaches the critical point  along the phase boundary.
At the critical point,   the shock height  is zero on the phase
boundary and it 
increases continuously as one proceeds 
away from the phase-boundary further into the 
shock phase.  In order to visualize  these features, 
 it is beneficial to obtain   the full solution for the 
density profile along with its boundary-layer.  Boundary 
layer analysis is useful for this purpose  since it 
allows us to generate a uniform approximation
 for solving the steady-state particle density equation 
across the entire lattice. 
This steady-state equation can be obtained from the large time- and  
length-scale limit (hydrodynamic limit)  
 of the statistically  averaged 
master equation that describes the particle dynamics in the 
discrete form.  For the simple 
exclusion case, it is possible to  obtain  
an  analytical solution of the steady-state hydrodynamic 
equation for the entire density profile. This, however, may not possible 
for more complex  ASEPs.

A fixed-point analysis of the hydrodynamic equation turns
 \cite{smfixedpt} out to be  
general and useful, 
 since it does not involve an explicit solution of the 
steady-state hydrodynamic 
 equation. In   
particle conserving models, a  boundary-layer saturates to   
 the constant bulk density profile  asymptotically. 
As a consequence of this, it is expected that the fixed-points 
of the boundary-layer equation  match with  the bulk density values. 
In other words, a boundary-layer, which is a  solution 
of the boundary-layer equation, is a  part of the 
 flow trajectory of the equation  flowing to 
the appropriate fixed-point on the phase plane. 
Thus, in order to find 
out the values of the bulk-densities in different phases, 
it is sufficient to determine the physically acceptable 
fixed-points of the boundary-layer equation. 
As a result,
the number of possible bulk phases 
is given by the number of these  fixed-points. 
Applying this method 
to a specific particle conserving two-species ASEP \cite{smfixedpt}, 
 it is found that this system has three distinct bulk phases 
corresponding to three fixed-points of the boundary-layer  equations.
In addition, it is possible to predict the nature of phase transitions,
locations of the boundary-layers {\it etc.} for this system. 
All these   predictions  
 match well with the results from numerical simulations \cite{popkov}.

In a particle non-conserving case, the density is not constant 
in the bulk, and therefore, the fixed-points of the boundary-layer 
do not provide 
the full profile since the details of the bulk dynamics 
is not considered in this approach.  However, it is still useful 
to obtain the  fixed-points of the boundary-layer equations 
along with  their stability properties  in order to 
predict the  possible shapes of the  density profiles under different 
boundary conditions.  
In the present paper, we consider a particle number non-conserving model 
where particles interact repulsively. Our  aim is to extend the 
fixed-point analysis  to a system  with  non-constant bulk density. 
We show how this analysis helps us predict 
possible shapes of the density profiles under different boundary 
conditions and also understand the properties of different kinds of   
shocks present in the density profile.
This particular model  is chosen because of certain nontrivial shapes 
of the density profiles with different kinds of shocks. 

The plan of the paper is as follows. In the following section, we 
describe the model. This section also contains  brief discussions on the  
hydrodynamic approach,  boundary-layer analysis and some of the known
results. In section III, we present the  phase-plane analysis of the 
boundary-layer equations for the present model. 
There are separate subsections on  the  
boundary-layer equation, its  fixed-points, stability analysis of the 
fixed-points and possible shapes of shocks. Section IV presents the
predictions of the possible shapes of the density profile under 
different boundary 
conditions. In section V, we mention a few  general rules for 
predicting the shapes of the density profiles and some special 
features related to the shocks of this model.   
We end the paper with a summary   
in section VI.  
   
\section{Model}
\subsection{Discrete description}
The  asymmetric simple exclusion process that we consider here 
consists of a one-dimensional lattice of  $N$  
sites with lattice spacing, $a$.
 Particles are injected at  $i=1$  with rate $\alpha$ and 
withdrawn at $i=N$ at a rate $1-\gamma$. 
Particles, obeying mutual exclusion, hop to the right 
with rates that depend on the occupancy of the neighboring site as  
\begin{eqnarray}
1100\rightarrow 1010 \ \ {\rm at} \ \ {\rm a} \ \ 
{\rm rate} \ \ 1+\epsilon,\\
0101\rightarrow 0011 \ \ {\rm at} \ \ {\rm a} \ \ {\rm rate} 
\ \ 1-\epsilon,\\  
0100 \rightarrow 0010 \ \ {\rm at} \ \ {\rm rate} \ \ 1\\
1101 \rightarrow 1010 \ \ {\rm at} \ \ {\rm rate} \ \ 1.
\end{eqnarray}
Here,  $0<\epsilon<1$ and   $1$ ($0$) represents an occupied (unoccupied) 
site. For $\epsilon \neq 0$, there is an effective 
repulsion between the particles \cite{kls,hager,rakos}.
In addition, the  number of particles  is not conserved due to  
particle detachment, $1\rightarrow 0$,
 at a rate $\omega_d$ and attachment, 
$0\rightarrow 1$,
 at a rate $\omega_a$ at any site on the lattice. 
Particle attachment and detachment  are 
equilibrium like processes that do not give rise 
to any particle current.

\subsection{Hydrodynamic Approach and a brief description of  the
boundary-layer analysis}
The hydrodynamic approach is based 
on the lattice continuity equation  which equates   
the time evolution of 
the particle occupancy  at a given site  
with the difference of currents 
across its two neighboring bonds.  In the continuum description, the 
continuous time and space variables are $t$ and $x$ with the 
latter replacing, for example, the $i$th site as $i\rightarrow x=ia$. 
Upon doing a Taylor expansion of the statistically averaged 
continuum version of the lattice 
continuity equation in small $a$, one has the following  
hydrodynamic equation
\begin{eqnarray}
\frac{\partial \rho}{\partial t}+\frac{\partial J}{\partial x}+S_0=0,
\label{fulleqn}
\end{eqnarray}
for the averaged particle density $\rho(x,t)$. 
This equation has already been supplemented with the 
particle non-conserving  parts 
\begin{eqnarray}
S_0=-\Omega(\rho_L-\rho),
\end{eqnarray}
where  $\rho_L=\frac{\omega_a}{\omega_a+\omega_d}$ and 
$\Omega=(\omega_a+\omega_d)N$. 
The current, $J(\rho)$,  consists of a bulk current 
$j(\rho)$ and a diffusive current  proportional to 
$\frac{\partial\rho}{\partial x}$ as
\begin{eqnarray}
J=-\epsilon_0\frac{\partial \rho}{\partial x}+j(\rho).\label{bigj}
\end{eqnarray}
Here, $\epsilon_0$ is a small parameter proportional to 
 $a$. The diffusive current part arises naturally  as one retains terms 
up to $O(a^2)$ in the Taylor expansion. 
In order to determine the  particle density, $\rho(x)$, 
in the steady-state ($\frac{\partial \rho}{\partial t}=0$), 
one has to solve the differential equation with appropriate boundary 
conditions. 
We consider the lattice-ends  to be attached to the 
particle-reservoirs which maintain constant densities 
$\rho(x=0)=\alpha$ and $\rho(x=1)=\gamma$. 
The diffusive current part  is crucial here  
since due to its presence, the hydrodynamic equation becomes  
a second order differential equation 
and as a result we can obtain a smooth solution 
satisfying both the boundary conditions.

$\epsilon=0$ is the usual ASEP  with only the exclusion 
interaction.  In this case, the  current density relation,    
$j(\rho)=\rho(1-\rho)$ is an exact one. 
The symmetric shape of the current 
about its maximum at $\rho=1/2$ is a consequence of its invariance 
under particle-hole exchange $\rho\rightarrow 1-\rho$.
 It is well understood  that the phase diagram 
has  low-density ($\rho<1/2$ at the bulk), high-density ($\rho>1/2$ 
at the bulk) and maximum current ($\rho=1/2$ at the bulk) phases
\cite{straley}. 
The particle-hole symmetry is retained in $\epsilon \neq 0$ models 
although the current changes non-trivially. At $\epsilon=1$, i.e. for 
  the extreme 
repulsion case, hops such as $0101\rightarrow 0011$  are forbidden. The
current, therefore, 
vanishes exactly at the  half-filling ($\rho=1/2$) with  the maximum  
current  appearing symmetrically  for densities  on the 
two sides of $\rho=1/2$. The exact form of the current 
 as a function of $\rho$ for arbitrary $\epsilon$  
can be found using  a transfer 
matrix approach \cite{hager} and it evolves from a single to a 
symmetric 
double peak structure as $\epsilon$ grows beyond $\epsilon_J\approx .8$. 
A simple, analytically tractable  form of the current  with double
peaks can be obtained by doing a double expansion of  the exact
current  about $\epsilon=\epsilon_J$  and $\rho=1/2$ \cite{jaya}. 
This leads to a quadratic form for the current 
\begin{eqnarray}
j(\rho)=(2r+u)/16-\frac{r}{2}(\rho-1/2)^2-u(\rho-1/2)^4,\label{jr}
\end{eqnarray}
where the constant term is chosen in such a way that $j(\rho)=0$
for $\rho=0, \ {\rm or} \  1$. We recover the non-interacting limit,
$j(\rho)=\rho(1-\rho)$ for $r=2$ and $u=0$. The double peak
shape appears for $r<0$. In the entire analysis below, 
we consider $r$ to be a small negative parameter and  $u>0$.

\begin{figure}[htbp]
  \begin{center}
   \includegraphics[width=3.5 in, clip]{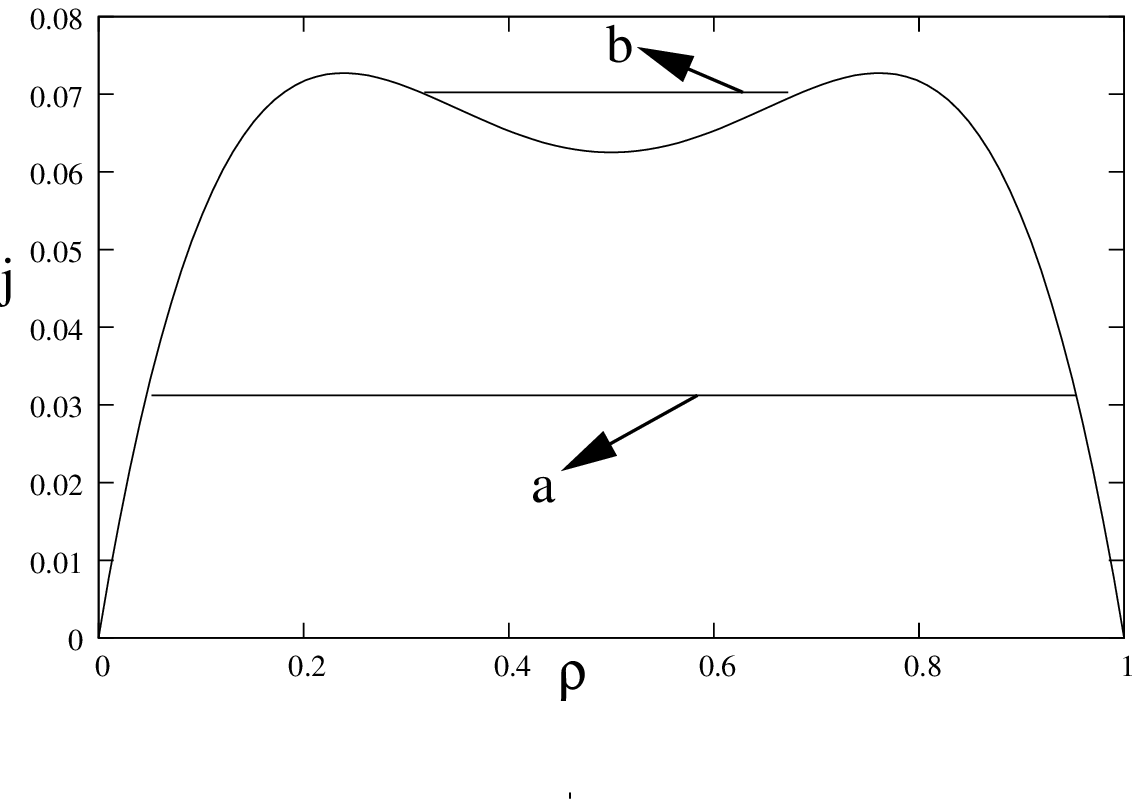}
    \caption{Current $j$ is plotted as a function of $\rho$. Lines 'a'
and 'b' represent an upward and a downward shock respectively.}
\label{fig:jvsrho}
  \end{center}
\end{figure}

For the 
 boundary-layer analysis, it is important to consider the bulk part 
and the narrow boundary-layer or the shock  
regions  of the density profile separately.    
These boundary-layers or shocks are formed over a narrow region of width
 $O(\epsilon_0)$ and they merge 
 to the bulk density in the appropriate asymptotic 
limit. In order to study the boundary-layer and its asymptotic
approach  to the bulk, 
one can rescale the position variable in (\ref{fulleqn})
 as $\tilde x=(x-x_0)/\epsilon_0$,
where $x_0$ is the location of the center of the boundary-layer. 
Hence, for a  boundary-layer  satisfying the 
boundary condition at $x=1$, we 
have $x_0\approx 1$. For small $\epsilon_0$, the boundary-layer
 approaches the bulk density 
 in the $\tilde x\rightarrow -\infty$ 
limit and satisfies the boundary condition at $\tilde x=0$. 
In terms of $\tilde x$, the steady-state 
hydrodynamic equation   is 
\begin{eqnarray}
 \frac{\partial^2 \rho }{\partial \tilde x^2}-
\frac{\partial j}{\partial \tilde x}-
\epsilon_0 S_0=0.\label{blayer}
\end{eqnarray}
Since $\epsilon_0$ is a small parameter, the effect of the particle
non-conserving term, $S_0$,  on the boundary-layer
 is negligible. As a result the
total current, $J=j(\rho)-\frac{\partial \rho}{\partial \tilde x}$
is constant  across the boundary-layer. A shock, therefore,
 can be represented  by a horizontal line connecting two
 densities in the $j-\rho$ plane as shown in figure (\ref{fig:jvsrho}).
For an   upward shock ($\frac{\partial\rho}{\partial \tilde x}>0$), 
this line lies below the $j(\rho)$ curve and the reverse is true
for a downward shock ($\frac{\partial\rho}{\partial \tilde x}<0$). 
As a result,  while for $r>0$, only upward
shocks are possible,  for $r<0$, there can be
 density profiles with a  downward shock and
double shocks. Double shocks can be represented 
by two horizontal lines on the $j-\rho$ plane below 
 the two peaks in $j(\rho)$.

 To zeroth order in 
$\epsilon_0$, the final boundary-layer equation is  
\begin{eqnarray}
 \frac{\partial^2 \rho }{\partial \tilde x^2}-
\frac{\partial j}{\partial \tilde x}=0.\label{blayerfinal}
\end{eqnarray}
In the boundary-layer language, the  solution of this equation 
is known as the inner solution. 
To obtain the bulk part of the density profile, one can ignore the 
diffusive current part in $J$ for  small $\epsilon_0$. 
The steady-state equation that gives the bulk part 
of the density profile is    
\begin{eqnarray}
\frac{dj}{dx}+S_0=0.\label{outer}
\end{eqnarray}
The solution of this equation  for the bulk part of the profile 
is known as the outer solution. These  inner and outer solutions
contain several integration constants  
which are fixed by the boundary conditions and other 
matching conditions of the boundary-layer and the  bulk  under  
various limits. Since the  slope of the outer solution 
is obtained from
\begin{eqnarray}
\frac{d\rho}{dx}=-S_0/\frac{dj}{d\rho},\label{outer}
\end{eqnarray}
for a given $\rho$, the slope  
depends crucially on the signs  of $(\rho_L-\rho)$ and 
$\frac{dj}{d\rho}$. For all the analysis below, we 
consider $\rho_L$ to be large and 
$\alpha$ and $\gamma$ to be much smaller than $\rho_L$.

\subsection{Known results }
\label{known}
Double peak structure of the current-density relation leads to  
two maximum current and one minimum current phases in 
  the phase diagram of  the particle conserving  repulsion model
\cite{hager}. 
In the maximum and  minimum current phases, the  bulk density values 
 are those  at which the current attains its  maximum and minimum
values respectively. 
 With these new phases,  the phase diagram for 
this   model becomes more 
complex than its non-interacting counterpart.

Combining the techniques of boundary-layer analysis  and the 
results from numerical solutions, the   phase diagram has been
obtained for the particle non-conserving repulsion  model \cite{jaya}.
The phase diagram has a lot
of interesting features including a tricritical point at $r=0$. In the
$\alpha-\gamma-r$ phase diagram, this is a special point where two critical
lines meet. It has been found that three different phase diagrams are 
possible for $r>0$, $r=0$ and $r<0$. For $r>0$, the current-density 
plot is  symmetric around $\rho=1/2$ with a maximum at $\rho=1/2$. 
The nature of the phase diagram is qualitatively similar to the 
mutually exclusive 
case with one single critical point. For $r<0$, with a double 
peak structure of the current-density plot, the phase diagram is
more complex with more than one critical point and 
three different shock phases with the density profile having a
 single upward shock,  double upward shocks  and 
one upward and one downward shock \cite{jaya,rakos}.

The low-density peak can give rise 
to a low-density upward shock ($\rho(x)\le 1/2$ in the shock part)
in the density profile. 
A single  shock of this kind  
can be represented by a  horizontal line in the $j-\rho$ plane 
below the low-density peak. The critical 
point corresponds to a situation where the horizontal line  reaches the 
peak position implying a shock of zero height. 
The second distinct critical point that 
involves both the peaks of the current-density plot is not 
symmetrically related to this. The density profile, here, 
has two upward shocks, in which one  is a low-density shock and 
one is a high-density shock  with $\rho>1/2$. 
The low-density shock, in this case, has the maximum height
with its high-density end saturating to $\rho=1/2$.
The high-density shock which is due to the high-density  peak 
of the current-density plot can be of varying height. 
The critical point corresponds to the special point 
where this high-density  shock 
has zero height. In addition to these regions,
 there are regions in the phase diagram, where density profiles 
with a  downward shock or a single, symmetric upward shock are   
found. 

In view of  the symmetry of the $j-\rho$ diagram,  it  is natural to 
expect the two critical points to be related through this symmetry.
Previous work, however, shows that the shapes of the 
density profiles are not  related through this symmetry 
 near these two special points.
Unlike the low-density shock, 
 the high-density shock in the density profile is always 
accompanied by a low-density shock of maximum height. 
The following 
analysis clearly reveals the reasons behind such asymmetries.

\section{ Phase-plane analysis of the boundary-layer equations}
In the following subsections, we determine the fixed-points of the 
boundary-layer equation and their   stability properties. 
These fixed-points are the special points to which the 
boundary-layer solution saturates in the appropriate limit. The knowledge 
about the fixed-points and their stabilities can, therefore, be used to 
our advantage to find out, for example, 
 the bulk densities to which a shock or a boundary-layer
 saturates at its two edges. 

\subsection{Boundary-layer equation}
Substituting  the expression for  $j(\rho)$ as given in  (\ref{jr})
and integrating the  boundary-layer equation, (\ref{blayerfinal}) 
once, we have 
\begin{eqnarray}
\frac{d\rho_1}{d\tilde x}+\frac{r}{4}\rho_1^2+
\frac{u}{8}\rho_1^4=C_0.
\label{finalinner}
\end{eqnarray}
Here $\rho_1=2\rho-1$ and 
$C_0$ is the integration constant. 
The saturation of the boundary-layer to the 
bulk density, $\rho_{1b}$,  is ensured 
by choosing the integration constant as 
\begin{eqnarray}
C_0=\frac{r}{4}\rho_{1b}^2+\frac{u}{8}\rho_{1b}^4.\label{c0eqn}   
\end{eqnarray}
As per equation (\ref{jr}), $C_0$ is related to the excess current 
(positive, negative or zero) measured from $\rho=1/2$ (half-filled 
case). The entire 
analysis in the following is done in terms of $\rho_1$ for which   
the boundary conditions are  $\rho_1(x=0)=\alpha_1=2\alpha -1$ and 
$\rho_1(x=1)=\gamma_1=2\gamma-1$.  

\subsection{fixed-points}
\begin{figure}[htbp]
  \begin{center}
   \includegraphics[width=3.5in,clip]{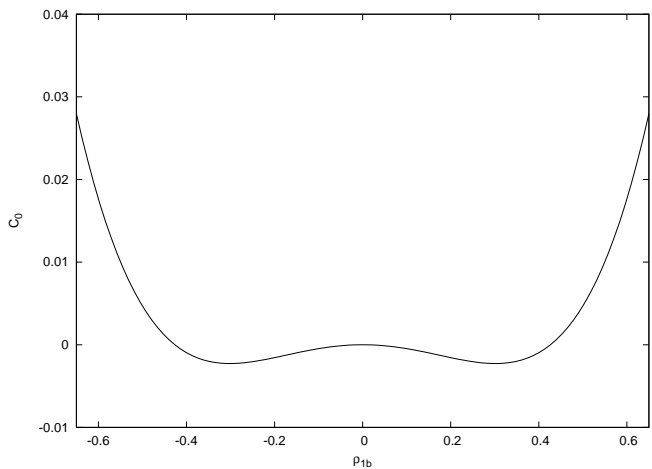}
    \caption{$C_0$ plotted as a function of $\rho_{1b}$ with 
$r=-.2$ and $u=2.2$ }
\label{fig:doublewell}
  \end{center}
\end{figure}

$C_0$ can be plotted for various $\rho_{1b}$ varying from 
$-1$ to $1$. For $r<0$, $C_0$ has a symmetric double well structure 
around $\rho_{1b}=0$ (see figure (\ref{fig:doublewell})).
The fixed-points, $\rho_1^*$,
of  equation (\ref{finalinner}), are the solutions of the 
algebraic equation  
\begin{eqnarray}
\frac{u}{8}\rho_1^4+\frac{r}{4} \rho_1^2-C_0=0.
\end{eqnarray}
In general, there are four possible solutions for the fixed-point 
as \begin{eqnarray}
\rho_1^*=\pm[\frac{\mid r\mid \pm\sqrt{r^2+8 C_0 u}}{u}]^{1/2}.
\label{fixed}
\end{eqnarray}
The value of $C_0$ depends on $\rho_{1b}$,  the bulk density 
to which the boundary-layer solution  saturates.
As a consequence, for a given $C_0$,  the corresponding $\rho_{1b}$
is always a fixed-point. For the same $C_0$, there are, however, 
 other fixed-points
which are determined from (\ref{fixed}).  Hence, from the 
information about one saturation density $\rho_{1b}$, the other 
saturation density of the shock can always be determined. 
The approach to various  fixed-points has to be, of course, consistent 
with their  stability properties.  These stability properties 
of various fixed-points are discussed in the following subsection. 

If $C_0$ is positive, there can be only two real 
fixed-points of opposite signs. The positive and negative 
fixed-points denoted respectively  as $\rho_{1\pm}^*$ are
\begin{eqnarray}
\rho_{1\pm}^*=\pm[\frac{\mid r\mid +\sqrt{r^2+8 C_0 u}}{u}]^{1/2}.
\end{eqnarray}
If $C_0<0$, there are 
four fixed-points for $\rho_{1b}$. In all these cases, the 
 fixed-points are symmetrically 
located on the either side of the origin. The positive $\rho_1$ 
fixed-points are 
\begin{eqnarray}
\rho_{1,2+}^*= [\frac{\mid r\mid \pm\sqrt{r^2-8 \mid C_0\mid u}}{u}]^{1/2}
\end{eqnarray}
and the negative $\rho_1$ fixed-points are 
\begin{eqnarray}
\rho_{1,2-}^*= -[\frac{\mid r\mid \pm\sqrt{r^2-8 \mid C_0\mid u}}{u}]^{1/2}.
\end{eqnarray}
Here, the subscripts $1$ and $2$ 
correspond to the $+$ and  $-$ signs inside the square bracket respectively. 
It is important to notice that for  $C_0<0$, all the fixed-points 
become imaginary when $\mid C_0\mid>\frac{r^2}{8u}$. 
 As $C_0$ approaches 
this  lowest negative value, the pair of fixed-points on 
the positive and 
negative  sides  approach each other and they  merge  at 
$C_0=-\frac{\mid r\mid^2}{8u}$. At this special value, the fixed-points 
are $\rho_{1m}^{*\pm}=\pm(\frac{\mid r\mid}{u})^{1/2}$. 
For $C_0=0$, there are three fixed-points, $\rho_1^*=0$ and 
$\rho_{10}^{*\pm}=\pm (\frac{2\mid r\mid}{u})^{1/2}$.

Numerical values of the fixed-points for some special values of $C_0$ with 
$r=-.2$ and $u=2.2$ are mentioned below. 
For $C_0=0$,  the nonzero fixed-points are  
$\rho_{10}^{*\pm}=0,\ \ \pm.426$.
For these values of $r$ and $u$, no real fixed-points are present if 
$C_0<-\frac{r^2}{8u}=-.00227$. 
 At this special value of $C_0$, the two  fixed-points  
are $\rho_{1m}^{*\pm}=\pm .30151$.

\subsection{Stability analysis of the fixed-points}
For $C_0>0$, a linearization of equation (\ref{finalinner}) around the 
fixed-points with $\rho_1=\rho_1^*+\delta\rho_1$  
leads to the following stability equation
\begin{eqnarray}
\frac{d\delta \rho_1}{d\tilde x} =
\frac{-\sqrt{\mid r\mid^2+8 u C_0}}{2}\rho_1^*\delta\rho_1.
\end{eqnarray}
This  implies that the fixed-points 
$\rho_{1+}^*$  and $\rho_{1-}^*$
are, respectively,  stable and  unstable.

\begin{figure}[htbp]
  \begin{center}
   \includegraphics[width=3in,clip]{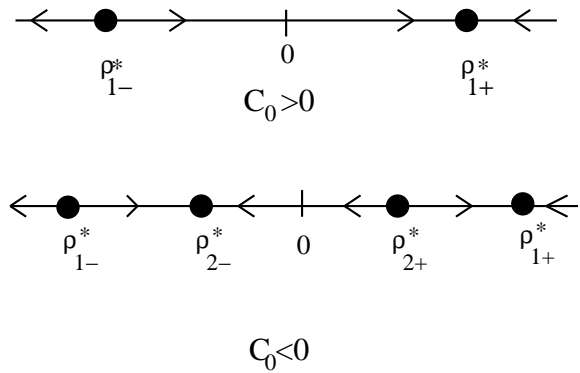}
    \caption{flow behavior of the fixed-points }
\label{fig:flow}
  \end{center}
\end{figure}

Similarly, for  $C_0<0$, the general  stability equation is 
\begin{eqnarray}
\frac{d\delta\rho_1}{d\tilde x}
=\frac{\rho_1^*}{2} \ \delta\rho_1 ({\mid r\mid}-
{u} {\rho_1^*}^2).
\end{eqnarray} 
The flow around the fixed-points can be obtained by 
substituting the explicit expressions  of the fixed-points. 
Figure  (\ref{fig:flow}) shows the stability properties 
of various  fixed-points for $C_0>0$ and $C_0<0$.

\begin{figure}[htbp]
  \begin{center}
   \includegraphics[width=3.5in,totalheight=3.5in,clip]{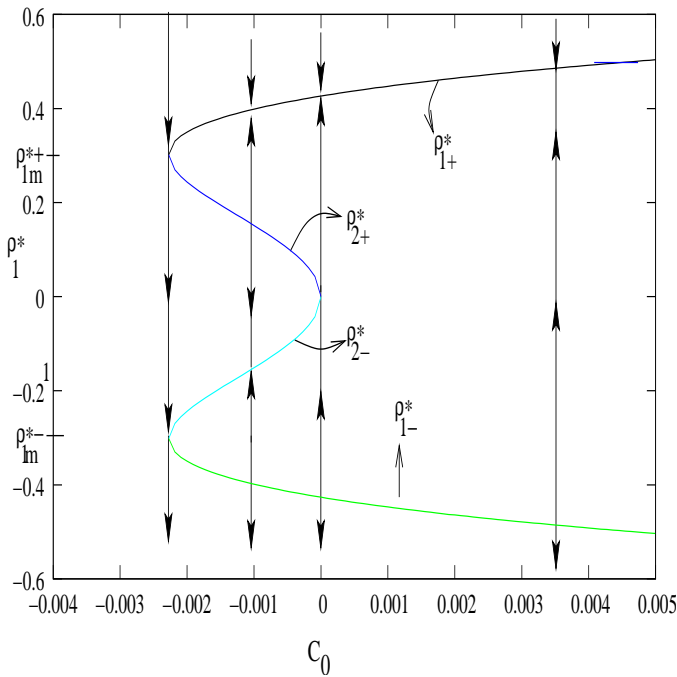}
    \caption{Fixed-points are plotted for different values of $C_0$ with
$r=-.2$ and $u=2.2$. $\rho_{1\pm}^*$ and  $\rho_{2\pm}^*$ are different 
fixed-points as mentioned in the text.
The vertical solid  lines with arrows show the flow behavior  
of the fixed-points.  }
\label{fig:c0route}
  \end{center}
\end{figure}

The stability property of the $\rho_1^*=0$ fixed-point for $C_0=0$ and  
for the pair of fixed-points for $C_0=-\frac{r^2}{8u}$ cannot be 
determined from the linear analysis. However, the flow around the 
fixed-points can be predicted from the  continuity of the 
flow behavior  as $C_0$ approaches these special values.
Fixed-points, their stability properties and how the fixed-points 
change with $C_0$ are combinedly shown in figure  (\ref{fig:c0route}).

\subsection{Shocks for various $C_0$}
Since $C_0$ can be expressed purely in terms of  $J$,
 with its constant parts subtracted,
it remains constant across a shock or a boundary-layer.  
In principle, using  equation (\ref{c0eqn}), 
one can  obtain the  
value of  $C_0$ along the continuously varying parts 
of the density profile. Hence,
as we move along a  density profile having bulk shocks,  
$C_0$ changes as per equation (\ref{c0eqn}) along the outer solution 
parts of the  profile with intermediate constant values 
across the shock or inner solution  regions.  
The value of $C_0$ in 
the shock region is  fixed by one of the bulk density 
values to which the shock saturates.   


With the information on the 
 possible values of $C_0$ in the entire 
range of $\rho_{1b}$ and, hence,   the knowledge about the 
corresponding  fixed-points,  it is possible to list the 
kind of shocks that can be observed.
 
 Let us assume  that the shocks or the boundary-layers 
approach the bulk densities $\rho_{1r}$ or $\rho_{1l}$ 
as $\tilde x\rightarrow \pm \infty$ respectively. Since $\rho_{1l}$ and 
$\rho_{1r}$ are various fixed-points of the inner equation, the approach 
to these fixed-points has to  be consistent with the flow properties. 
A shock is called an  upward shock if  $\rho_{1l}<\rho_{1r}$. 
The reverse i.e. $\rho_{1l}>\rho_{1r}$ is true for a downward shock.

(i) $C_0>0$: In this case, there are two fixed-points $\rho_{1+}^*$ and 
$\rho_{1-}^*$ symmetrically located  around $\rho_1=0$ with   $\rho_{1-}^*$  
being an  unstable fixed-point. Thus if a shock is formed with $C_0>0$, it 
should  be an upward shock which  
approaches the fixed-points $\rho_{1r}=\rho_{1+}^*$ 
and $\rho_{1l}=\rho_{1-}^*$ as $\tilde x\rightarrow \infty$ and  
$-\infty$ respectively. 
The  shock height, in this case, is $\rho_{1+}^*-\rho_{1-}^*$.

(ii) $C_0<0$: In this case, four fixed-points lead to different kinds of 
  shocks.  

(a) It is possible to see a downward shock
with   $\rho_{1l}=\rho_{2+}^*$ and $\rho_{1r}=\rho_{2-}^*$.
The downward shock is thus symmetric around $\rho_1=0$. The flow  
in figure (\ref{fig:flow}) shows that a downward 
shock cannot involve  other fixed-points since that  would not 
be consistent with the stability criteria of the fixed-points.    

(b) There can be  small upward shocks  which lie entirely 
in the range  $\rho_1>0$.  We have already  referred these shocks  
 as high-density shocks. In terms of the fixed-points, the left and 
right saturation densities of the shock are 
$\rho_{1l}=\rho_{2+}^*$ and $\rho_{1r}=\rho_{1+}^*$, respectively.

(c) The third possibility is that of an upward shock entirely in 
$\rho_1\le 0$ range. Such a shock  has been referred  as a  
low-density shock. For this shock,  $\rho_{1l}=\rho_{1-}^*$ and 
$\rho_{1r}=\rho_{2-}^*$.  

(iii) $C_0=0$: There can be an upward shock with $\rho_{1r}=0$ and 
$\rho_{1l}=\rho_{1-}^*$. There can also 
be an upward shock connecting the densities $\rho_{1l}=0$ and 
$\rho_{1r}=\rho_{1+}^*$. These two shocks together appear as a 
large shock, symmetric around $\rho_1=0$.

Alternatively, different kinds of shocks can  tell us the range 
of values for $C_0$.

\section{Predictions about the shapes of the density profiles}

\begin{figure}[htbp]
  \begin{center}
  (a) \includegraphics[width=.35\textwidth,clip,
angle=0]{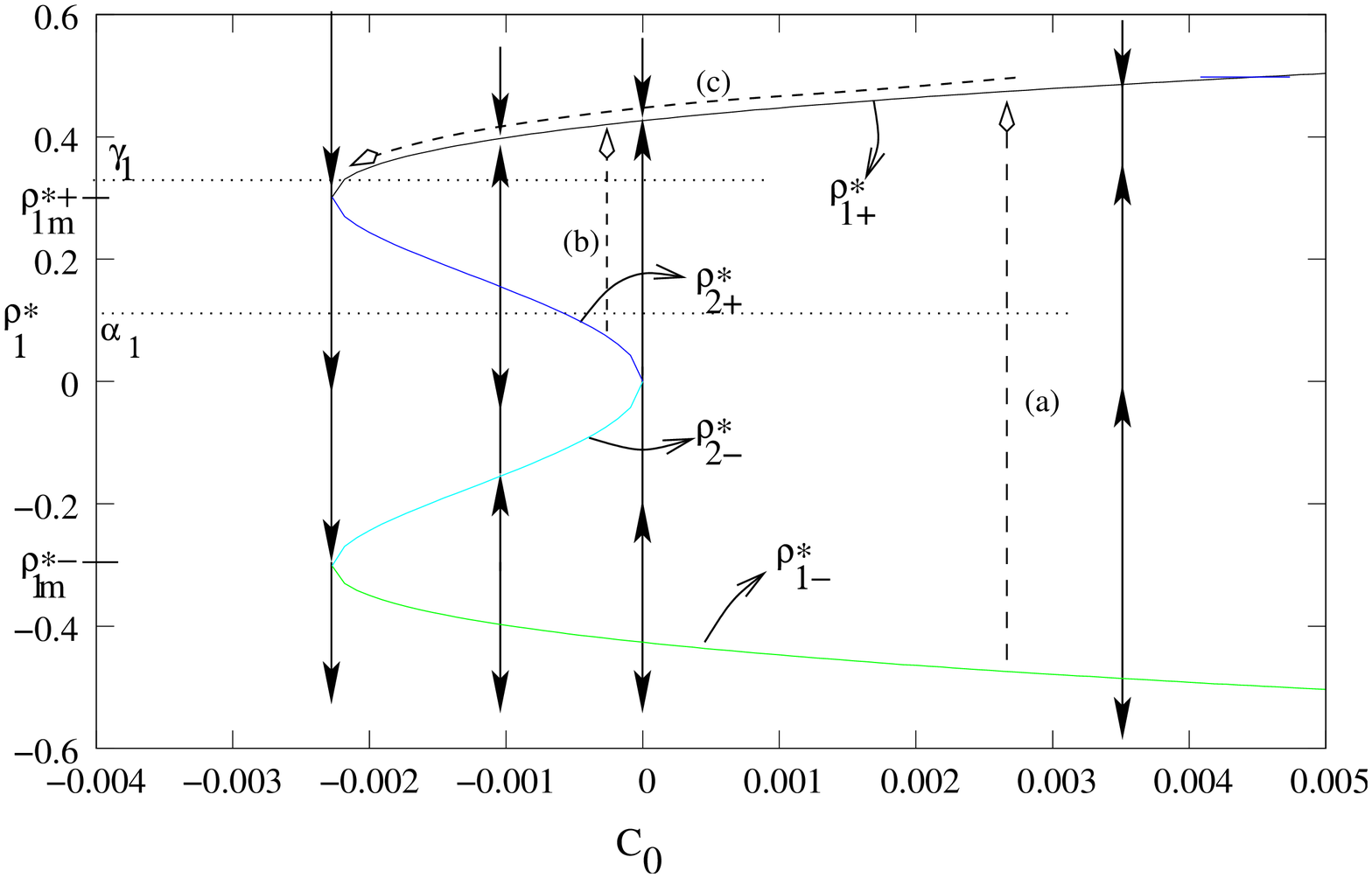}\\

(b)   \includegraphics[width=.4\textwidth,clip,
angle=0]{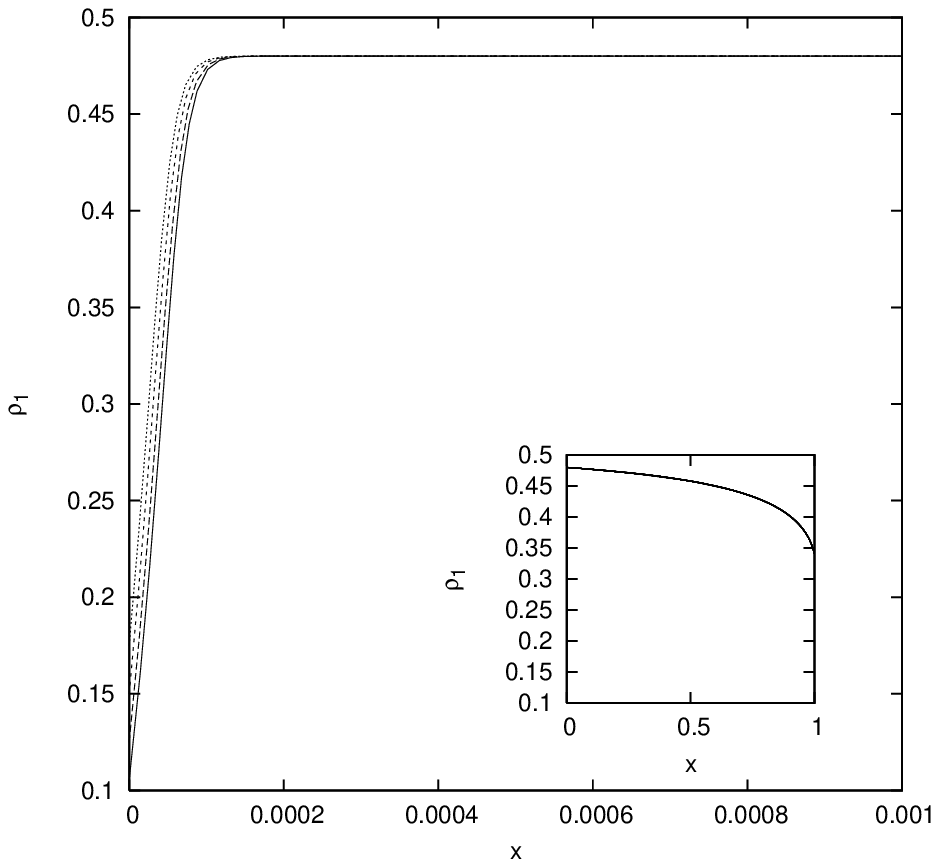}
    \caption{(a) Possible variations of the density, as one moves along the 
density profile from $x=0$ end, are shown on the $C_0-\rho_1^*$ plane.
 Bold dashed lines or curves  with open arrows show the variation 
of the density. Open arrows point in the direction of increasing $x$.
 Dotted lines mark the boundary conditions 
$\alpha_1$ and $\gamma_1$. Solid lines with arrows are the flow trajectories 
of the fixed-points. (b)  Numerical solutions for the density $\rho_1$ 
for various $\alpha_1$  with $\gamma_1=0.34$. 
The inset gives a zoomed view of the particle-depleted 
boundary-layer at $x=0$. No boundary-layer is 
formed near $x=1$.}
\label{fig:densprof1}
  \end{center}
\end{figure}

Based on figure (\ref{fig:c0route}),  we attempt to predict possible 
shapes of the density profiles for given boundary conditions 
$\alpha_1$ and $\gamma_1$. We consider only a few pairs of boundary conditions 
and based on this, we make certain general predictions in the next section.
The basic strategy for drawing the density profile is as follows. 
We first need to mark   $\alpha_1$ and $\gamma_1$ on the $\rho_1^*$ axis
of $C_0-\rho_1^*$ plane.
Starting with either of the boundary conditions, we change $\rho_1$, 
along the curve in figure (\ref{fig:c0route}), in 
a way that we reach  the other boundary condition in the end of our move. 
While doing so, we may allow a discontinuous variation of $\rho_1$ 
along a vertical constant-$C_0$ line,
 provided it does not violate the flow property.  Such a  discontinuous 
change in  $\rho_1$  appears in the form of a shock or a boundary-layer in 
the density profile. The  dashed, vertical lines in figure 
(\ref{fig:densprof1}a), for example, are the constant-$C_0$ lines along 
which the density may change.  
Such a dashed-line, therefore, 
represents a   boundary-layer  or a shock  in the  density profile. 
Two densities at which a shock or a boundary-layer saturates,  
are those at which a  particular,  constant-$C_0$ line, 
representing a shock or a boundary-layer, intersects the curves.
 This method, however, sometimes leaves us with 
different options for the density profile. All these possibilities are 
shown on $C_0-\rho_1^*$ plane  for each pair of boundary conditions
 individually.

\begin{figure*}[htbp]
  \begin{center}
(a)   \includegraphics[width=.35\textwidth,height=3in,clip,
angle=0]{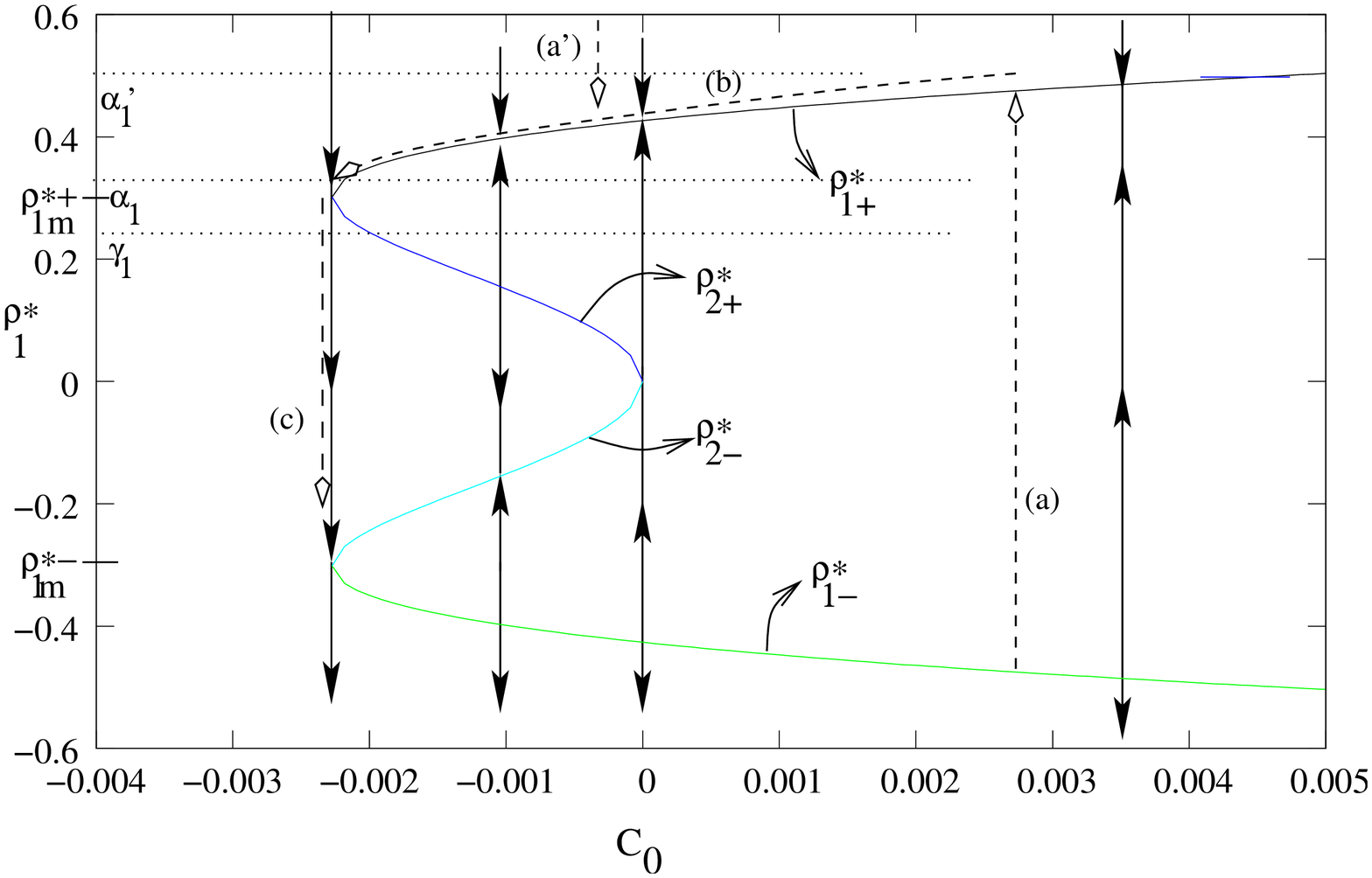}
(b)   \includegraphics[width=.37\textwidth,clip,
angle=0]{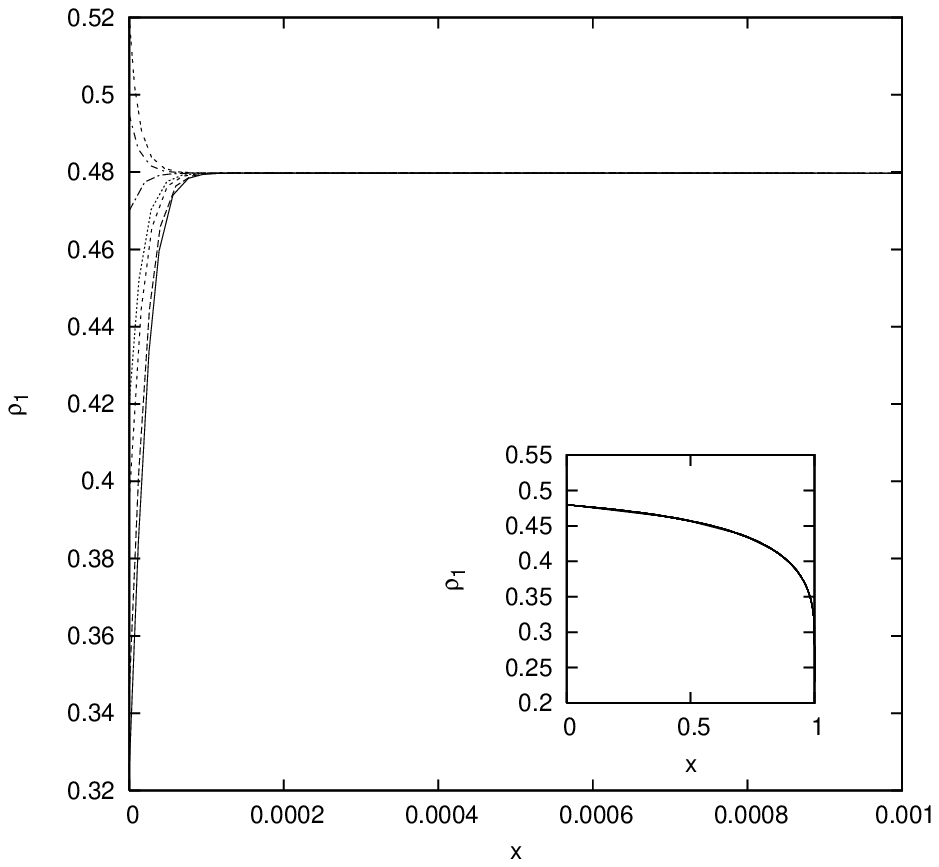}\\
(c)   \includegraphics[width=.37\textwidth,clip,
angle=0]{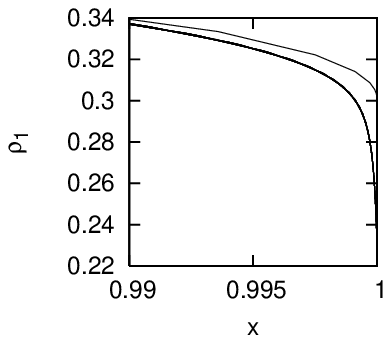}
    \caption{(a) Same captions as figure (\ref{fig:densprof1}a)  
except that here we explicitly 
show possible density variations for two different left boundary 
conditions specified by 
$\alpha_1$ and $\alpha_1'$.
(b) Numerical solutions for   $\rho_1(x)$
for various $\alpha_1$  with $\gamma_1=0.23$. The main figure provides
 a zoomed view of the boundary-layers near $x=0$. 
The entire density profile over the entire lattice is shown
 in the inset.
(c) A zoomed view of  the same  density profiles  near the $x=1$. 
This shows the particle depleted boundary-layers near $x=1$.}
\label{fig:densprof2}
  \end{center}
\end{figure*}

\subsection {Density profiles with only boundary-layers} 
Suppose we consider a situation where $\alpha_1,\ \gamma_1>0$
with $\alpha_1<\rho_{1m}^{*+}$ 
and $\gamma_1>\rho_{1m}^{*+}$. 
There can be a  possibility where  the density profile has 
a particle-depleted  boundary-layer 
($\frac{d\rho_1}{dx}\mid_{x=0}>0$) at $x=0$  
satisfying the boundary conditions  $\rho_1(x=0)=\alpha_1$. 
This boundary-layer can  be  represented by a vertical 
line similar to (a) in figure (\ref{fig:densprof1}a). This is consistent 
with the  flow property  
that suggests the approach of the boundary-layer  to the  fixed-point   
$\rho_{1+}^*$ as $\tilde x\rightarrow \infty$. On the other hand, 
in the $\tilde x\rightarrow -\infty$ limit, which corresponds 
to  the unphysical 
negative  $x$ region, the boundary-layer  saturates  to the 
unstable fixed-point $\rho_{1-}^*$. After the boundary-layer, the 
density may decrease continuously along (c) on the $\rho_{1+}^*$ branch
and satisfies the boundary condition,
 $\rho_1(x=1)=\gamma_1$. 
There can be another possibility 
where the particle-depleted boundary-layer at $x=0$ is represented by  
a vertical line similar to 
(b) joining the fixed-points  $\rho_{2+}^*$ and  $\rho_{1+}^*$. 
The boundary condition at $x=1$ is again satisfied by a decreasing 
density part parallel to (c). 
For this to be 
possible the condition $\gamma_1<\rho_{10}^{*+}$ is required. 
These two possibilities are distinct due to distinctly different 
values of $C_0$. This shows  the crucial role played by $C_0$ 
in deciding the density profile.  
Numerical solutions of the full steady-state hydrodynamic equation
presented   in figure (\ref{fig:densprof1}b) show the 
boundary-layers  saturating to a  bulk density  $\rho_{1b}\approx .48$. 
This implies  that the boundary-layers are indeed represented by  
(a) type vertical lines.

\begin{figure}[htbp]
  \begin{center}
 (a)  \includegraphics[width=.4\textwidth,clip,
angle=0]{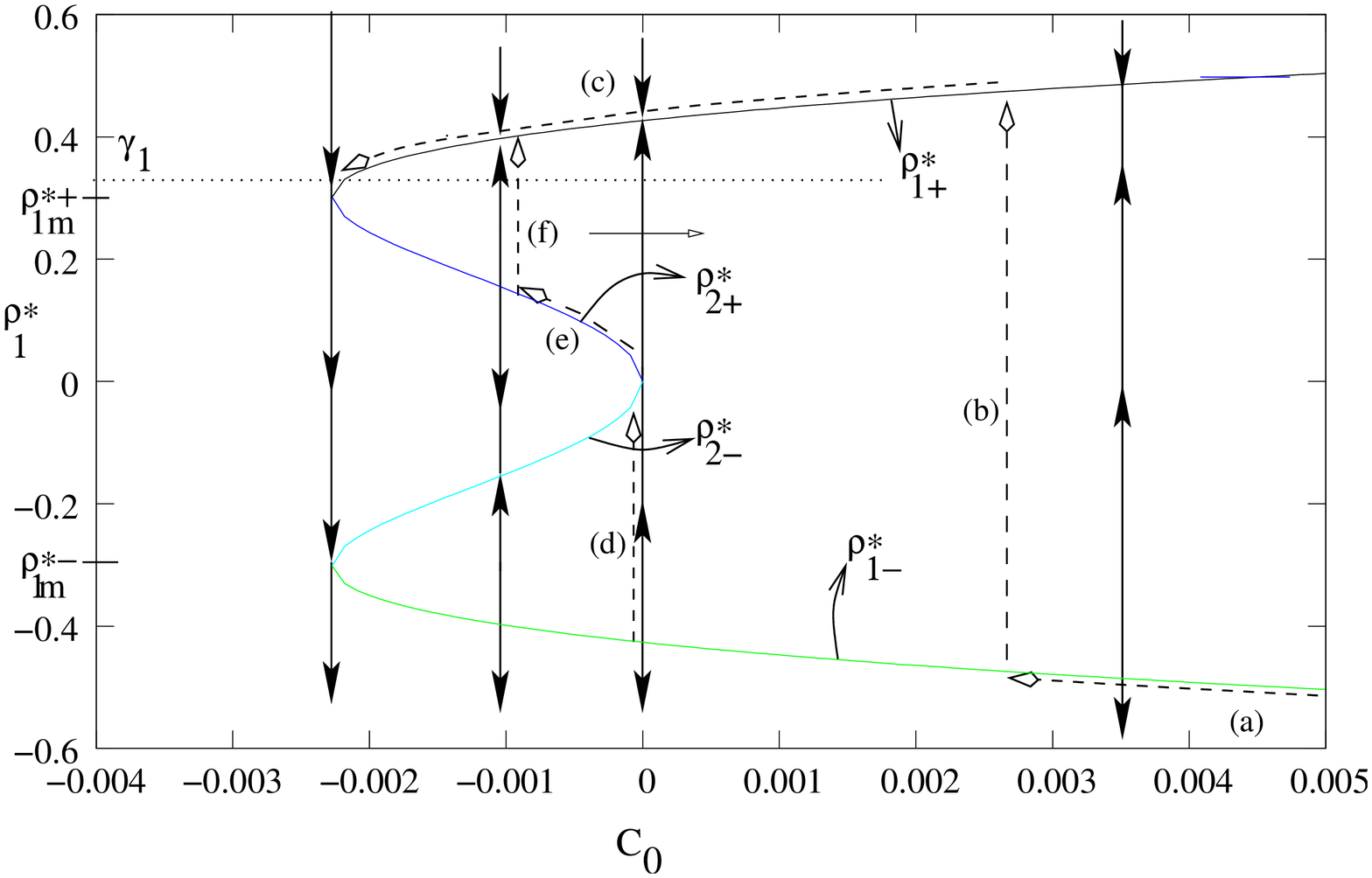}\\
 (b)  \includegraphics[width=.4\textwidth,clip,
angle=0]{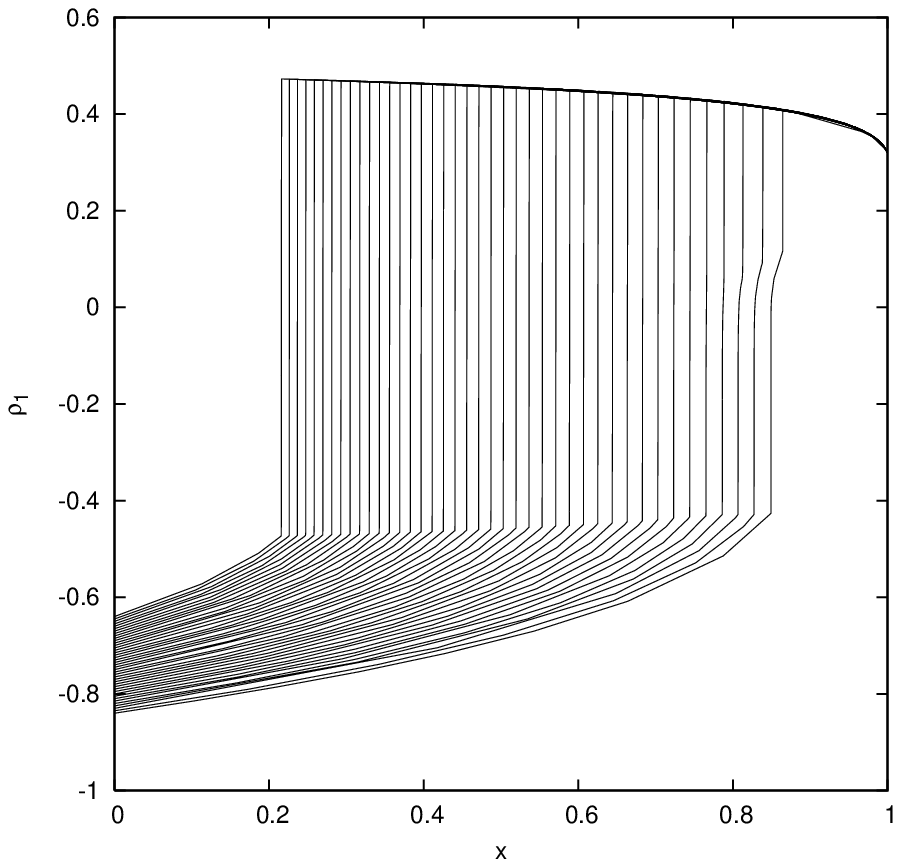}
    \caption{ (a) This figure has the same caption as figure 
(\ref{fig:c0route}).
(b) Plot of the density profiles  $\rho_1(x)$
for various large negative values of $\alpha_1$  with $\gamma_1=.32$. 
No boundary-layer is formed at $x=0$ or $x=1$. A few  density profiles 
 towards right have  double shocks.}
\label{fig:densprof3}
  \end{center}
\end{figure}

(2) Next, we consider $\alpha_1,\gamma_1>0$ with 
$\alpha_1>\rho_{1m}^{*+}$ and $\gamma_1<\rho_{1m}^{*+}$. 
In this case too, the  density profile can satisfy the 
boundary condition 
at $x=0$ through a boundary-layer 
that can be represented by a line similar 
to (a) in figure (\ref{fig:densprof2}a).
This would be a particle depleted boundary-layer at $x=0$. 
In order to satisfy the other boundary condition, 
the density should decrease till $\rho_{1m}^{*+}$ along path (b) on
 the $\rho_{1+}^*$ branch and then satisfy the right boundary condition 
 through a particle-depleted boundary-layer 
along (c) in figure (\ref{fig:densprof2}a). 
The boundary condition at $x=0$ can also be satisfied by vertical 
lines coming from above the $\rho_{1+}^*$ branch leading to particle-rich 
boundary-layers 
($\frac{d\rho_1}{dx}\mid_{x=0}<0$) at $x=0$. 
These lines could be (a') type lines 
in figure (\ref{fig:densprof2}a) satisfying the boundary condition 
 $\alpha_1'$.   
Both particle-depleted and particle-rich boundary-layers are present 
in the density profiles of   figure (\ref{fig:densprof2}b) obtained 
by solving the full steady-state hydrodynamic equation numerically.

\subsection{Density profiles with upward shocks}
 With $\gamma_1>\rho_{1m}^{*+}$ 
   and $\alpha_1$ large negative, possible
shapes of the
density profile can be  of the following kinds. We start with
$\rho_1(x=0)=\alpha_1$. $\rho_1$ decreases continuously 
along $\rho_{1-}^*$ branch along 
the dashed line (a) in figure (\ref{fig:densprof3}a). 
After this part,  an upward  shock, represented 
by  a vertical line similar to either (b) or  (d) appears.
In the latter case, the dashed line (a) should be extended further 
till it reaches the low-density end of (d).  
If the shock is represented by (b), it is a 
large shock, symmetric around $\rho_1=0$. 
In the second case, the shock is a low-density shock with 
the saturation densities being
$\rho_{1r}=0$ and $\rho_{1l}=\rho_{10}^{*-}$. 
If the density approaches the 
$\rho_{1+}^*$ branch after the large shock, the boundary condition 
 at $x=1$  can be 
satisfied after that by  a  decrease in density along (c) on 
this branch. If the shock 
is of  (d) kind, the density has to change further 
to satisfy the  right boundary 
condition. Upon reaching   $\rho_1=0$ value, the density may 
change along  $\rho_{2-}^*$ or the 
$\rho_{2+}^*$ branch. The flow around $\rho_1^*=0$, however, suggests
that the density variation only  along 
$\rho_{2+}^*$ branch (path (e) in the figure (\ref{fig:densprof3})) 
is possible. The continuously   increasing part  
along (e) 
 is then  followed by another upward shock, given by  line (f),
 taking the density to $\rho_{1+}^*$ branch. 
The boundary condition is then satisfied by a continuously 
decreasing part along a  (c) type line.
Numerical solutions in figure (\ref{fig:densprof3}b) are 
consistent with these predictions.

\subsection{Density profiles with downward  shocks}
\begin{figure}[htbp]
  \begin{center}
(a) \includegraphics[width=.4\textwidth,clip,
angle=0]{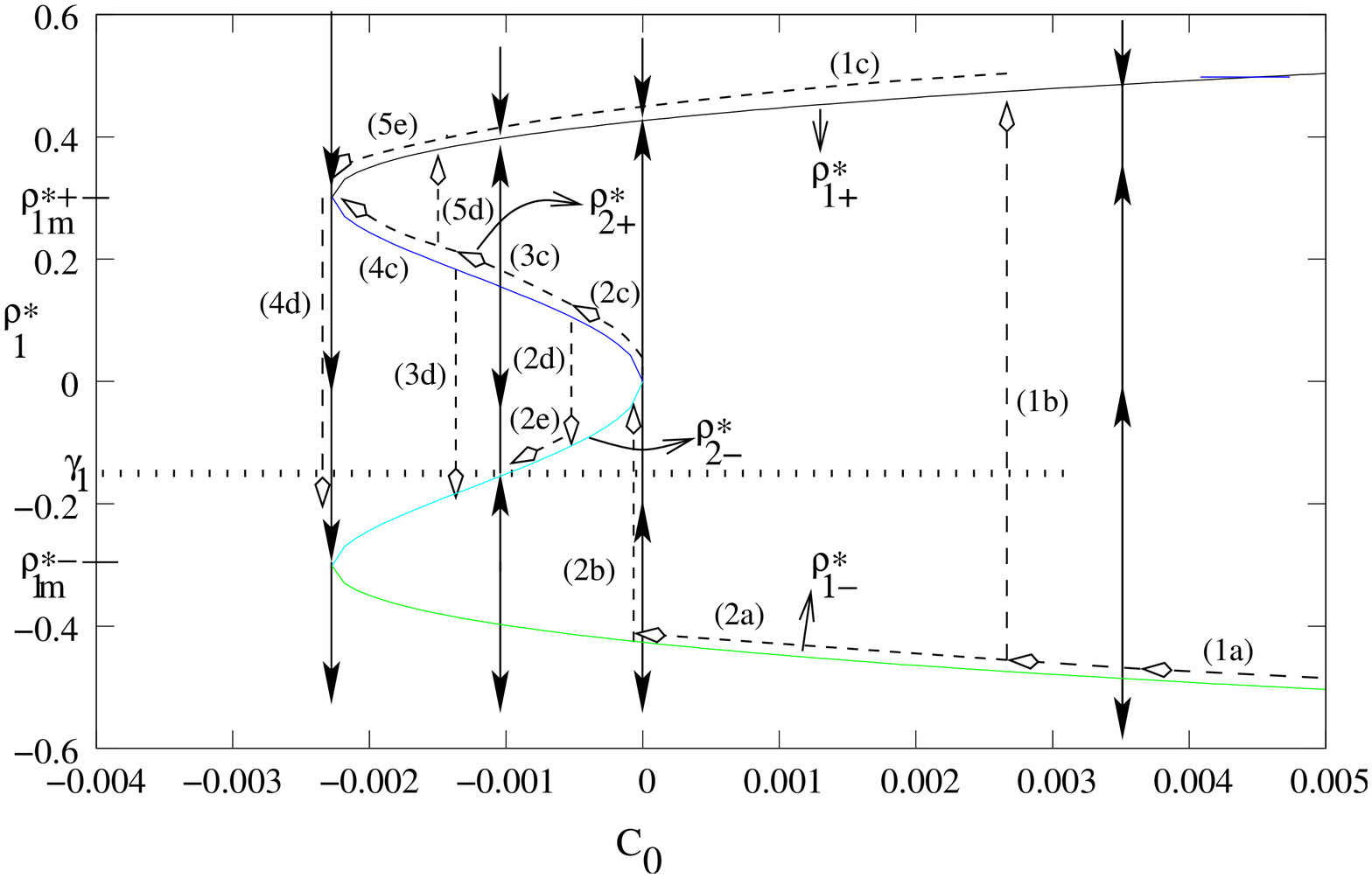}\\

 (b)  \includegraphics[width=.4\textwidth,clip,
angle=0]{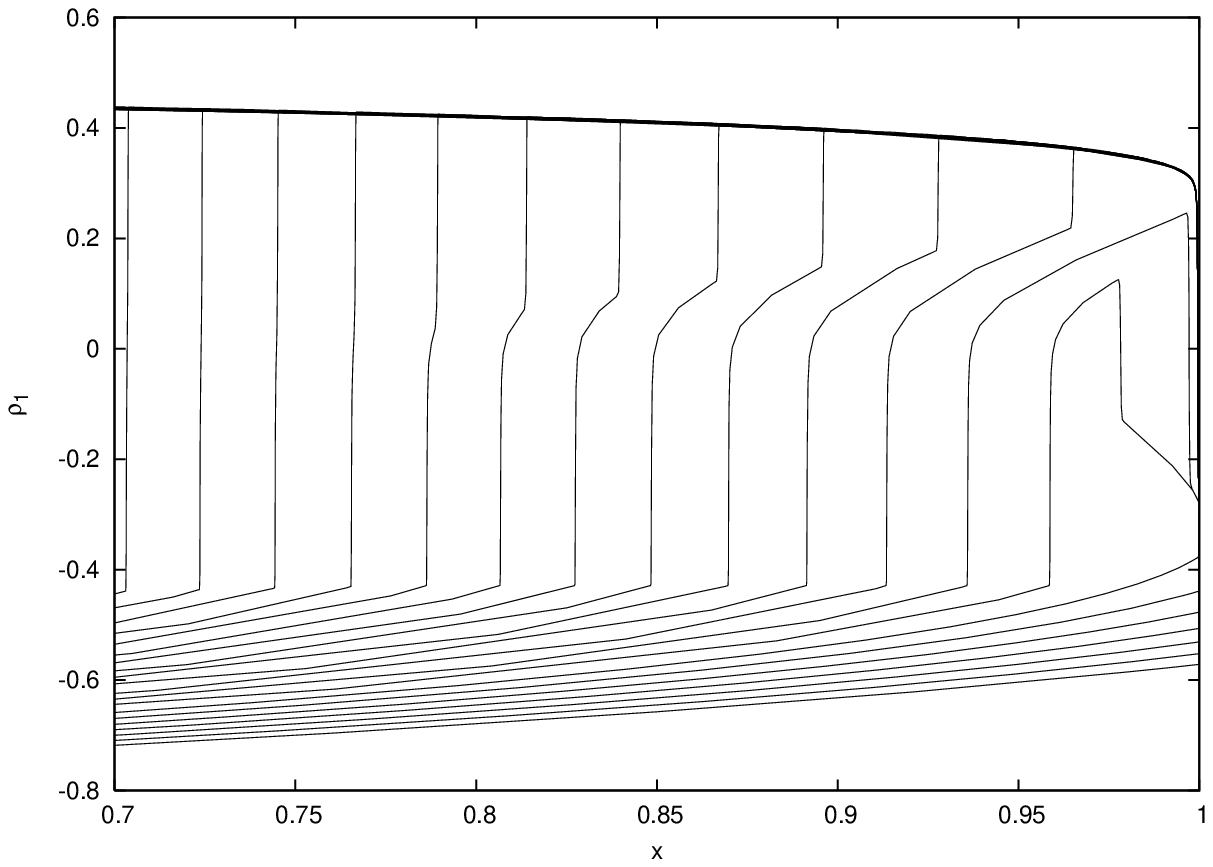}\\

 \includegraphics[width=.5\textwidth,clip,
angle=0]{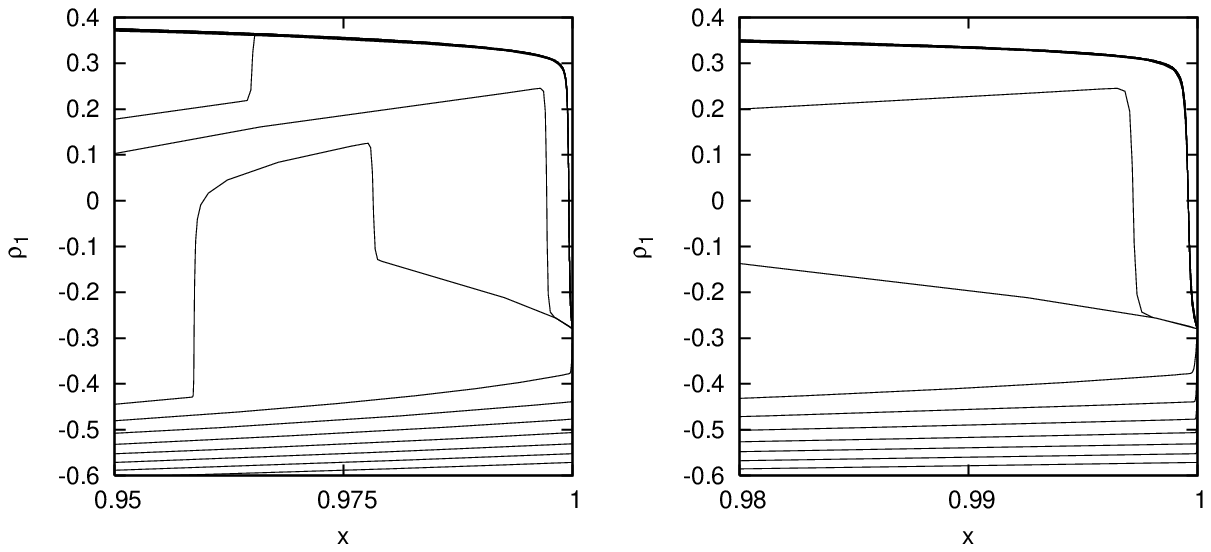}(c)
    \caption{
(a) The trajectory of the density on $C_0-\rho_1$ plane. Five possible 
trajectories are shown. These trajectories are distinguished by 
different numbers. Alphabetical sequence represents the variation of the 
density along increasing $x$. 
(b) Plot of the density profiles  $\rho_1(x)$
for various $\alpha_1$  with $\gamma_1=-.28$. (c) Zoomed views of the 
of the boundary-layers of the density profiles in (a) near $x=1$. }
\label{fig:densprof4}
  \end{center}
\end{figure}

We next  consider a case $\rho_{1m}^{*-}<\gamma_1<0$ 
with $\alpha_1$ increasing from large negative values.
Here we specifically mention how the density profile changes 
as $\alpha_1$ is changed keeping $\gamma_1$ fixed. In the process, 
we observe how a density profile with a downward shock appears. 
Let us assume that 
our starting $\alpha_1$  lies somewhere on the $\rho_{1-}^*$ branch.  
$\rho_1$ increases from $\rho_1(x=0)=\alpha_1$ along $\rho_{1-}^*$ 
branch till it reaches the boundary $x=1$. This continuously 
increasing part is represented by (1a) in figure 
(\ref{fig:densprof4}a).
The density, then, satisfies the 
right  boundary condition through a boundary-layer which can be, for 
example, represented by a vertical line like (1b), in figure 
(\ref{fig:densprof4}a). 
This line takes the solution  from the unstable fixed-point 
$\rho_{1-}^*$  to the  stable fixed-point  $\rho_{1+}^*$. 
Since the vertical line (1b)  
passes through $\gamma_1$ before reaching the $\rho_{1+}^*$ branch, the 
boundary-layer satisfies the boundary condition 
 before saturating to the 
positive fixed-point, $\rho_{1+}^*$. 
The boundary-layer at $x=1$ is, therefore, a part 
of this vertical, constant-$C_0$ line.  

As $\alpha_1$ is increased slightly, the route of  
the density along $\rho_{1-}^*$ branch  remains the same but 
this time the density reaches a higher value than the previous 
$\alpha_1$ case  before increasing sharply 
as a  boundary-layer satisfying 
the boundary condition at $x=1$. 
 As $\alpha_1$ is increased further,  
for a given $\alpha_1$, the continuously increasing 
 part of the profile reaches  the low-density end of the line (2b).
After this, there is a shock in the density profile of (2b) kind. 
This is a low-density shock that takes the density to $\rho_1^*=0$.
 If this jump 
is near  the boundary, this shock becomes actually a boundary-layer 
that can help the density satisfy the boundary condition at $x=1$.
However, if this discontinuity is in the bulk, it 
is an upward  low-density shock. 
In case of a shock in the bulk, the density increases further 
along $\rho_{2+}^*$ branch (path (2c) in  figure 
(\ref{fig:densprof4})). The 
boundary condition, however,  demands a decrease in $\rho_1$. This is 
possible through a downward vertical line (similar to path (2d)) 
and then a continuously decreasing part (2e) along $\rho_{2-}^*$ branch. 
The path (2d) is a downward shock that is seen in figure 
(\ref{fig:densprof4}b) and (\ref{fig:densprof4} c).  
In case the density varies along lines (3c) and (3d), we have 
a downward boundary-layer 
near $x=1$. These possibilities are expected if $\alpha_1$ 
is increased further from its value that leads  to (2c) and (2d) type 
variations. 
As before, the  principle is that if the 
(3d) type  vertical line intersects  $\gamma_1$ line before reaching the 
$\rho_{2-}^*$ branch,  (3d) type line  represents a boundary-layer 
satisfying the boundary condition
 at $x=1$. If the reverse happens,  this downward vertical line 
  represents a downward 
shock at the bulk which needs to be followed by a continuously 
decreasing density  part
 along the $\rho_{2-}^*$ branch. This is a general principle which can be 
applied to other cases also to see the deconfinement of a boundary-layer 
giving rise to a shock in the bulk (see reference \cite{smsmb}).

With  further increase of $\alpha_1$, the density variation from 
from $x=0$ end, is still the same as before up to  
part  (3c) along the $\rho_{2+}^*$ curve except that 
 now  the density approaches closer 
to $\rho_{1m}^{*+}$ along (4c) like path. Finally for a given $\alpha_1$,
the density reaches the value  $\rho_{1m}^{*+}$.
After this, the boundary condition is 
satisfied through a depleted boundary-layer represented by vertical 
path (4d).
With further increase of $\alpha_1$, the density  cannot now go   
around $\rho_{1m}^{*+}$ to move to $\rho_{1+}^*$ branch 
due to the constraint from the stability property. 
In that case, the only option for the density is 
 to proceed along (4c)  but move to $\rho_{1+}^{*}$ branch along a 
vertical line  similar to (5d) before reaching the point $\rho_{1m}^{*+}$. 
There is now a second high-density upward shock in the 
density profile at larger $x$ (line (5d))
with the first shock being a low-density one represented by 
the line, (2b). 
 With the increase of $\alpha_1$, the (5d) type vertical  line 
moves to higher values of $C_0$. 
 The boundary condition  at $x=1$  is  now satisfied 
by the rest of the density profile where the 
density decreases continuously along  (5e)  
path to the minimum $\rho_{1m}^{*+}$ and after that decreases 
through  a depleted boundary-layer represented by the line (4d). 
Thus, for example, for this value of $\alpha_1$, we see the following 
parts in the density profile as we move 
along the density profile from its   $x=0$ end. 
(i) A continuously varying density profile that satisfies  the boundary 
condition $\rho(x=0)=\alpha_1$. 
(ii) This is followed by a low-density upward shock of maximum 
height connecting $\rho_{1l}=\rho_{10}^{*-}$ and $\rho_{1r}=0$.
(iii) Beyond this shock, there is again a continuously increasing part.
(iv) This is followed by an upward, high-density shock.
(v) Beyond this high-density shock, the density decreases  continuously 
 to $\rho_{1m}^{*+}$.
(vi) The last part is a particle-depleted boundary  layer, that saturates to 
$\rho_{1m}^{*+}$ for $x<1$ and  satisfies the boundary condition at 
$\rho_1(x=1)=\gamma_1$. All  these features can be verified from 
the density profile in  figure (\ref{fig:densprof4}a).

If $\alpha_1$ is increased further, the upward high-density  shock 
(vertical lines like (5d)) will move towards higher $C_0$ values. 
For certain $\alpha_1$, the low 
and the high-density shocks merge and there is  a big symmetric shock
with $C_0=0$. 
Beyond this $\alpha_1$, the big 
upward shock still persists and this is followed 
by a continuously varying part  (similar to (1c)) along which the density 
decreases and approaches $\rho_{1m}^{*+}$. The boundary condition is again 
satisfied by the particle-depleted boundary-layer represented by (4d).

\section{General predictions}
Since the  boundary-layers or shocks are the special features through 
  which the density profiles are distinguished,  
our general predictions are more on the kind of shocks or 
boundary-layers that can be seen under various boundary conditions. 

\subsection{Shocks and boundary-layers}
Downward shock in the bulk or a particle-depleted boundary-layer at $x=1$: 
Either of these features appears  whenever the 
density profile decreases through a jump discontinuity from the 
$\rho_{2+}^*$ branch to the $\rho_{2-}^*$ branch or from 
$\rho_{1m}^{*+}$ to $\rho_{1m}^{*-}$. The condition for this is  
$\gamma_1<\rho_{1m}^{*+}$. The value of $\alpha_1$ is somewhat 
flexible since it is possible to see these features both for $\alpha_1$
positive or negative.

Upward symmetric shock in the bulk or depleted boundary-layer at $x=0$: 
This is seen whenever the density profile
jumps from $\rho_{1-}^*$ to $\rho_{1+}^*$. This happens for various 
combinations of $\alpha_1$ and $\gamma_1$ such as $\alpha_1<0$ or 
$\alpha_1>0$ with  
$\gamma_1>\rho_{1m}^{*+}$ or $\gamma_1<\rho_{1m}^{*+}$. At $x=0$, the 
density profile may start with a continuously varying part followed by a 
symmetric large shock, or it can satisfy the boundary condition
 at $x=0$ with the help 
of a boundary-layer. In both cases, the discontinuity in the density 
 corresponds to a discontinuous jump from $\rho_{1-}^*$ 
to $\rho_{1+}^*$.  

 The presence or absence of a boundary-layer 
 at $x=0$ is specified completely by the value of $C_0$ at $x=0$. 
Let us assume that $\rho_1=\alpha_1$ line intersects the curve on figure 
(\ref{fig:c0route}) at a value $C_0=C_0(\alpha_1)$. A condition as  
$C_0(\alpha_1)=C_0(x=0)$, would mean a  continuously varying 
density near $x=0$. If these two values are unequal, it would imply 
the presence of a boundary-layer. For example, for 
 $\alpha_1>\rho_{1m}^{*+}$, a particle-rich or a particle-depleted 
 boundary-layer  appears   if $C_0(\alpha_1)>C_0(x=0)$ and 
$C_0(\alpha_1)<C_0(x=0)$, respectively. However, it is important to
pay attention to certain situations which are forbidden due to 
the stability properties. For example, if 
$\alpha_1<\rho_{1m}^{*-}$ , a boundary-layer  with 
$C_0(x=0)<C_0(\alpha_1)$ is not possible.

Double shock: In this case, the density profile has 
both high and low-density upward  shocks 
with the low-density shock having
  maximum possible height, $\rho_{10}^{*-}$. 
In order to have a high-density shock, the
 lower end of the  high-density shock must be   
on the $\rho_{2+}^{*}$ branch. 
 The density can reach this branch only via  
$\rho_1=0$ point. The only way the density can reach the $\rho_1=0$ 
 point is through a low-density shock 
represented by the $C_0=0$  line across the negative lobe. A low-density 
 shock representing a jump across the negative lobe along $C_0=0$ line 
has the maximum possible height. 

 A density-profile with double 
shock may appear for $\alpha_1<\rho_{1m}^{*-}$ and 
$\gamma_1<\rho_{1m}^{*+}$ or $\gamma_1>\rho_{1m}^{*+}$. 
 In case of $\gamma_1>\rho_{1m}^{*+}$, 
the density after the high-density shock varies continuously along 
 $\rho_{1+}^*$ branch to satisfy the boundary condition at $x=1$.
For $\gamma_1<\rho_{1m}^{*+}$, the second shock  
is possible for some $\alpha_1$. 
In this case, after reaching the $\rho_{1+}^*$ branch, the density 
decreases till $\rho_{1m}^{*+}$ and then decreases further  as a depleted 
boundary-layer at $x=1$ to satisfy the boundary condition.

It is interesting to note that 
although  it is possible to have a profile with only a low-density 
shock,  the same with a single   
high-density shock is never possible. The flow behavior suggests 
that a high-density shock has to be  always  accompanied by a
 low-density shock of maximum height.

Boundary-layer at $x=1$: As in the case of a boundary-layer at $x=1$, 
it is also possible 
to specify the conditions for a boundary-layer 
at $x=1$ by comparing the 
value of $C_0(x=1)$ with $C_0(\gamma_1)$. 
In general, a boundary-layer will appear at $x=1$ if these two values of 
$C_0$ are different. As an example, a downward boundary-layer for 
$\rho_{1m}^{*-}<\gamma_1<\rho_{1m}^{*+}$ appears if 
$C_0(x=1)<C_0(\gamma_1)$.

\subsection{Saturation of the shock}
 From equation (\ref{finalinner}), we find that near  
the saturation to a bulk density $\rho_{1b}$, the slope of the 
boundary-layer is given by 
\begin{eqnarray}
\frac{d\delta\rho_1}{d\tilde x}=(\frac{\mid r\mid }{2}-
\frac{u}{2}\rho_{1b}^2)\rho_{1b}
\delta\rho_1
\end{eqnarray}
where it is assumed that the   boundary-layer density 
 is $\delta\rho_1$ away 
from the saturation value, $\rho_{1b}$. 
This shows, that the saturation of the 
boundary-layer to the bulk is in general 
exponential except for three special points. 
The saturation is  of  power-law kind, if  $\rho_{1b}=0$ or 
$\rho_{1b}=\rho_{1m}^{*\pm}=\pm(\frac{r}{u})^{1/2}$. The length scale 
associated with  the exponential approach of the shock to the bulk density 
diverges as the bulk density approaches these special values. As 
discussed in subsection \ref{known}, the 
critical points, correspond to  special boundary conditions 
$(\alpha_c,\gamma_c)$ at 
which the shock height across the positive or negative lobe reduces 
to zero. Therefore, the 
 approach to the critical  point is associated  with the  
continuous vanishing of the shock height along with 
 the divergence of the  length scale over which the shock 
 saturates to the bulk.

\section{Summary}
Here, we have considered an asymmetric simple exclusion 
process of interacting particles on a finite, one dimensional lattice. 
These  particles  have mutual repulsion in addition  to the
exclusion interaction. Apart from the hopping dynamics of the particles,
the model also has particle attachment-detachment processes, 
which lead to particle non-conservation in the bulk. 
Such processes are known to exhibit boundary-induced phase transitions 
for which the tuning parameters are the boundary densities, 
 $\alpha$ and $\gamma$. 
In different phases, the average particle density distributions across 
the lattice     have distinct shapes with various types of discontinuous 
jumps from one density value to another. Here, we carry out  a phase-plane 
analysis for the boundary-layer
 differential equation to 
understand  how the fixed-points of the boundary-layer equation 
and their flow properties determine 
the shape of the entire density profile under  given boundary conditions. 
Such a fixed-point analysis has been extremely useful in 
understanding the phases and phase transitions of particle 
conserving models for which the constant bulk density values in different 
phases are given 
by the physically acceptable fixed-points of the boundary-layer equation. 
In addition, the number of 
steady-state phases, the nature of the  phase-transitions, the locations of 
the boundary-layers can be obtained analytically from the phase-plane 
analysis of  the boundary-layer equation. The  present work  
 provides  a generalization 
of the method to a particle non-conserving process.

To  apply this method, we  have considered 
the hydrodynamic limit of the statistically averaged 
master equation describing the particle  dynamics. The hydrodynamic 
 equation, describing the time evolution of the average particle 
density, looks like a continuity equation   
supplemented with the 
 particle non-conserving terms. The current 
contains the exactly known hopping current  and 
 a regularizing diffusive current part. 
 The boundary-layer equation,  which is the  
main focus of this work, can 
be obtained from the particle conserving part of 
the  hydrodynamic equation. For convenience, we use 
$\rho_1=2\rho-1$ for the boundary-layer equation. 
$\rho_1$ is related 
to the deviation from $\rho=1/2$ (half filled case). 

It is found that the  fixed-points, $\rho_1^*$, 
 of the boundary-layer  equation are determined in terms 
of a parameter $C_0$ related to the excess current measured 
from $\rho=1/2$ (half-filled case). 
 Since the fixed-points 
are dependent on $C_0$, one can plot the physically acceptable 
fixed-points as a function of $C_0$ on the $C_0-\rho_1^*$ plane. 
In the steady-state, 
the constancy of the current across a shock or a boundary-layer 
 implies that such objects  
can be represented by a fixed value of $C_0$.  
 The boundary-layers or shocks of the density 
profiles are represented by the constant-$C_0$ lines on this 
$C_0-\rho_1^*$ plot. The densities at which the  constant-$C_0$ 
line intersects the fixed-point branches 
are the densities to which the  
shock or the boundary-layer saturates. 
The discontinuous change of the density has to be consistent
with the stability properties  of the fixed-points. 
For given values of $\alpha$ and $\gamma$,  
we can start from the $x=0$ end of the density profile  and find out  
how  the density can  change along the profile 
as it proceeds to satisfy the boundary condition 
at $x=1$. 
This density variation  along the density profile 
can be conveniently marked on the  $C_0-\rho_1^*$ plot to 
see its consistency with the flow properties of the fixed-points.
 Our approach does not give any information about  the location of a 
shock  since  it does not 
involve the details of the bulk part of the profile. 
Instead, it is found that the   conserved quantity, $C_0$, 
 plays an important role in deciding the 
shape of the density profile.

The emphasis of our approach is on the boundary-layer equation which 
appears to  control the shape of the  entire density profile. Particle 
non-conserving processes are not important for the boundary layers.
This simplicity allows us not only to classify different kinds of 
density distributions,  but also to gain more physical insight 
as why some features of the  density profile are  evident under certain  
boundary conditions. Some of these features are mentioned in the 
list below. 

(a) When a density profile has two shocks, the low-density shock 
 is of maximum possible height. For given values 
of the interaction parameters, the height of the low-density 
shock can be obtained explicitly.

(b) It is possible to have a low-density shock alone in the profile 
but a high-density shock has to be always accompanied by a 
low-density shock of maximum height.

(c) A downward shock is produced by the deconfinement of a downward 
boundary-layer
at $x=1$. The condition on $\gamma$ for seeing a downward shock or 
a downward boundary-layer at $x=1$   can be 
precisely specified. 

(d) The symmetric two peak structure of the current as a function of 
the particle density is responsible for  a symmetric  two lobe 
structure of the fixed-points drawn on $C_0-\rho_1^*$ plane. 
The flow behavior of the fixed-points
around the two lobes are asymmetric. This is the  reason why 
the two critical points in the phase diagram  are not symmetrically 
related to each other. This asymmetry is reflected in 
the shapes of the density profiles  near these critical points. 
 
(e) For a given boundary condition, a density 
profile with only one  boundary-layer and no shock  can be fully 
specified by the value of $C_0$ at this end. 

In addition to these issues, this analysis also 
provides quantitative predictions regarding the heights of different 
kinds of shocks  and their  approach to the bulk  along with 
 the length scale associated with it.

{\bf Acknowledgement}
Financial support from the   Department of Science 
and Technology,  India   and warm hospitality of ICTP (Italy),
 where the work was initiated, are  gratefully acknowledged.

\end{document}